\documentclass[reprint,pra,superscriptaddress,showpacs,twocolumn]{revtex4-1}
\usepackage{graphicx}                           % Permite incluir figuras e grÃ¡ficos no formato .eps
\usepackage[ansinew]{inputenc}                  % Permite voce acentuar normalmente: Ã­ , Ã§ , Ã£ , ...
\usepackage{amsfonts}                           % Pacote para por simbolos matematicos especiais.
\usepackage[normalem]{ulem}                     % Pacote que permite sublinhar e/ou riscar palavras ou frases
\usepackage{amsmath}                            % Pacote que permite usar comandos especiaÃ­s nas fÃ³rmulas (como \dfrac).
\usepackage{lipsum}
\usepackage{mathrsfs}

\begin{document}
%\title{Two proposals to protect a qubit of the losses of an imperfect quantum memory using CQED techniques}
\title{Two proposals to protect a qubit using CQED techniques: inequality between atomic velocity dispersion 
and losses of a quantum memory}

\author{J. L. Santos}
\email{jefferson.santos@ufu.br}
\affiliation{Instituto de Física, Universidade Federal de Uberlândia, 38400-902, MG, Brazil}
\author{J. G. G. de Oliveira Jr.}
\email{jgojunior@uesc.br}
\affiliation{Departamento de Ci\^encias Exatas e Tecnol\'ogicas,
Universidade Estadual de Santa Cruz, 45.662--900, Ilh\'eus, BA, Brazil}
\author{J. G. Peixoto de Faria}
\email{jgpfaria@des.cefetmg.br}
\affiliation{Departamento de Física e Matemática, Centro Federal de Educação Tecnológica de Minas Gerais, 30510-000 Belo Horizonte, MG, Brazil}
\author{M. C. Nemes}
\email{Deceased.}
\affiliation{Departamento de Física, Instituto de Ciências Exatas, Universidade Federal de Minas Gerais, 30123-970 Belo Horizonte, MG, Brazil}

\date{\today}

\begin{abstract}
We present in this work an analysis of the damage imposed by the atom on the field state inside a 
lossy superconducting cavity. To access such effects, we propose two procedures to preserve a qubit 
of the decay effects of an imperfect quantum memory: the first by means of an quasi-instantaneous 
phase kick applied in the atom, and the second by means of controlled resonant and dispersive 
interactions. We immediately demonstrate that, in both procedures, the dwell time of the qubit in 
the cavity increases significantly, being expressively higher for the second. A relation between the 
inaccuracy of the preparation of the atomic beam and the quality of the cavity arises naturally from 
our calculations for each procedure. This result is unprecedented, and sets out the rules to increase 
the dwell time of the qubit. %Moreover, we show that the dispersion of the atomic velocity impose new 
%noises for the field state inside the cavity, that are similar to the action of a phase reservoir.
\end{abstract}

%\pacs{03.65.Ta,03.67.Mn,03.65.-w}
%\keywords{}

\maketitle
\section{Introduction}

Combined systems of atoms and photons, have been studied extensively to construct promising and efficient quantum networks for information processing and communication \cite{monroe,kuhn}. In these quantum networks, quantum state transfer between photons and atoms (matter), and storage of quantum information are of the utmost importance. Therefore, numerous methods to implement the quantum state transfer and quantum memory have been proposed and investigated in various manners. In context of cavity quantum electrodynamics (CQED), the investigation of a coherent atom-photon interaction in high quality microwave \cite{nobel} or optical \cite{reiserer} cavities, provides deep insights in fundamental quantum phenomena and in quantum information procedures. In these studies is essential to manipulate the atom-field system in a coherent and reversible unitary way, minimizing the adverse influence of nonunitary irreversible decoherence processes induced by environment. Maître et al. in \cite{maitre} have describe that, a qubit initially carried by a two-level atom can be transferred to the Electromagnetic Field (EMF) mode inside a cavity of high quality factor. Then a second atom, after a delay time $\tau$, collects the qubit stored in the EMF mode. In this work, the cavity functions as a quantum memory. But the quantum information stored in the cavity is quickly lost, due to the process of decoherence caused by the environment.

A good quantum memory should be able to keep the qubit preserving its coherence over a long period of time when compared to the dissipation imposed by the medium. But it is not only the environment that causes processes of decoherence, the atom when interacting with the field depending on the perfection of how it was prepared can impose new noises to the field state inside the cavity, provoking new processes of decoherence. Here we investigate the damages imposed by the atom on the field state inside a lossy superconducting cavity. To access such effects, we propose two procedures that allow to partially protect a qubit of the decoherence imposed by the losses of the cavity, the first procedure of preservation is by means of a quasi-instantaneous phase kick applied to the atom, while the second procedure is through interactions atom-field resonant and dispersive controlled. For this, we consider a Fock state (with 0 or 1 photon) into an imperfect cavity and an atomic beam with $N$ two-level atoms in the ground state ($\left| g\right\rangle$). The Fock state (qubit), originally in the cavity (quantum memory) is shared with the atomic beam through controlled interactions. We have shown that, in both procedures, we will have gains in the dwell time of the qubit, from this we did a detailed study of the influence that the dispersion of the atomic velocity exerts in the procedures of CQED. In addition, a study of the damping channels and their relations with the dispersion of the atomic velocity was carried out, in order to better understand the coherence loss suffered by the field inside $C$.

The paper is organized as follows. In Sec. II we discuss the physical system and we present our two preservation procedures, as well as the gains they provide. In Sec. III is made a detailed study of the damages that the dispersion of the atomic velocity imposes in the experiments of CQED, as well as the rules that must be satisfied to reach expressive gains in the dwell time of the qubit inside a cavity. Already in Sec. IV we observed the relationship between the dispersion of the atomic velocity and the damping phase channel. Finally, in Sec. V is calculated the fidelity in order to compare how much information is lost with the dispersion of the atomic velocity in relation to the passage of $N$ atoms through the cavity.

\section{Physical system}

\begin{figure}[h]
	\centering
	\includegraphics[scale=0.3]{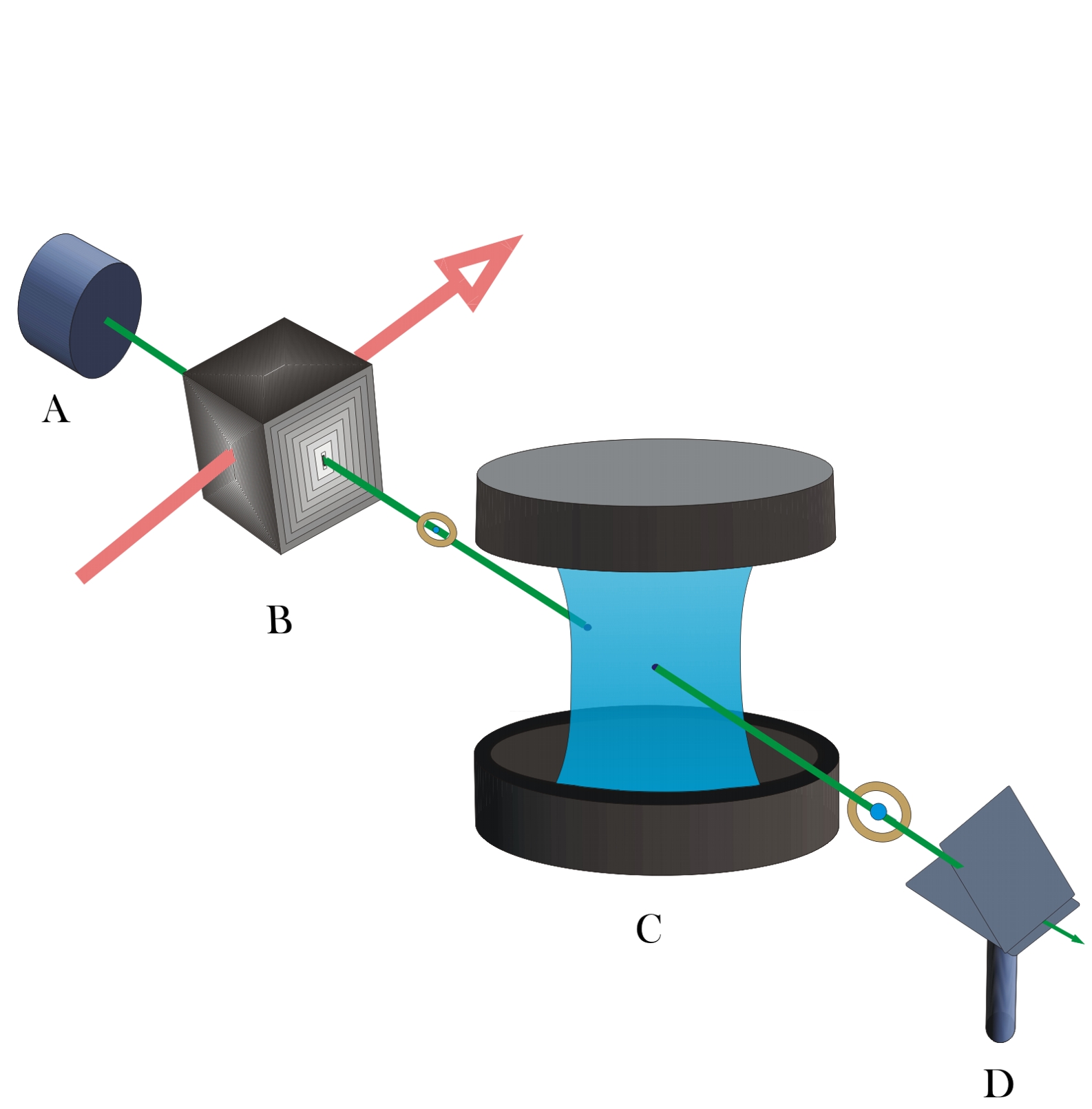}
	\caption{General scheme of the experimental apparatus used in CQED with Rydberg atoms.}
	\label{fig1}
\end{figure}

To explain our method, consider the experimental setup that is depicted in Fig. \ref{fig1}. An atomic beam effusing from oven $A$ ejects atoms (usually rubidium) that, when entering in box $B$, has its velocity selected and is prepared in circular Rydberg states \cite{raimond,nussenzveig}. The all atoms in the ground state $\left|g\right\rangle$ cross one high-finesse superconducting cavity $C$, where encounters a Fock state (with 0 or 1 photon). The superconducting cavity is in a Fabry-Perot type configuration, made of two spherical niobium mirrors with a Gaussian geometry and photon damping times of 130 $ms$ \cite{gleyzes2007}. The cavity is prepared at a low temperature ($\approx 0.6K$) to reduce the average number of thermal photons. After $C$ finally the atom can be measured in D in te ground $\left|g\right\rangle$ or excited $\left|e\right\rangle$ states. The full Hamiltonian this system can be written as
\begin{equation}
{H}={H}_0+{H}_I\ ,
\end{equation}
where
\begin{eqnarray*}
{H}_0&=&\frac{\hbar\omega_{eg}}{2} {\sigma}_z + \hbar\omega_c({a}^{\dagger}{a}+\mathbf{1}/2)\ ,\\
{H}_I&=&\hbar G({a}{\sigma}_++{a}^{\dagger}{\sigma}_-)\ .
\end{eqnarray*}
Here, $H_0$ describes the noninteracting system, where the cavity mode is associated with the annihilation ${a}$ and creation ${a}^\dagger$ operators, the atomic operator is given by ${\sigma}_z=\left|e\right\rangle\left\langle e\right| - \left|g\right\rangle\left\langle g\right|$, already $\omega_{eg}$ and $\omega_c$ are the frequencies of atomic transition and cavity mode, respectively. The term ${H}_I$ describes the atom-cavity interaction, which is the Jaynes-Cummings model \cite{jaynescummings} under the rotating wave approximation (RWA), the coupling constant $G=\Omega_0/2$, where $\Omega_0$ is the vacuum Rabi frequency and $\sigma_{\pm}$ are the atomic operators that describes the promotion and fall, respectively.

In the experiments of atom-field interaction, the frequencies of the cavity mode and atomic transition are of extreme importance, since they can be adjusted in order to the system interact resonantly ($\omega_{c}=\omega_{eg}$) or dispersivement ($\omega_{eg}-\omega_{c}=\delta$, which is known as detuning). In the resonant interaction there is an exchange excitation between the atom and the cavity mode, so that at the end of the interaction the atom-field system is entangled. On the other hand, in the dispersive interaction the probability of an exchange excitation occurring between the atom and the cavity mode is practically zero, so that at the end of the interaction the eigenvectors of the atom-field system are factored, i.e., the interaction does not entangle the system. The effective Hamiltonian that will generate this dispersive interaction is \cite{romero}
\begin{eqnarray}\nonumber
{H}_{ef}&=&\frac{\hbar\omega_{eg}}{2}{\sigma}_z+\hbar\omega_c({a}^{\dagger}{a}+\mathbf{1}/2)+\hbar\omega\big[({a}^{\dagger}{a}+\mathbf{1})\left|e\right\rangle\left\langle e\right|\\
& &-{a}^{\dagger}{a}\left|g\right\rangle \left\langle g\right|\big]\ ,
\end{eqnarray}
where $\omega\equiv G^2/\delta$.

The cavity $C$ has a high quality factor, however this quality factor is not infinite, which means the existence of losses. We can model this process of cavity losses by means of the dynamic of open quantum systems, which can be obtained by tracing over all of the reservoir degrees of freedom from the total system $\rho(t)=tr[\rho_{tot}(t)]$, where $\rho_{tot}(t)$ is the total density matrix of the system plus its reservoir. The exact master equation of the reduced density matrix $\rho(t)$ for the open system, in the Lindblad form
\begin{eqnarray}\nonumber
\frac{d}{dt}{\rho}(t)&=&\frac{1}{i\hbar}[{H},{\rho}(t)] +\kappa(1+\bar{n})\lbrace[{a}{\rho}(t),{a}^{\dagger}] + [{a},{\rho}(t){a}^{\dagger}]\rbrace\\ \label{eq.mestra}
& &+ \kappa\bar{n}\lbrace[{a}^{\dagger}{\rho}(t),{a}] + [{a}^{\dagger},{\rho}(t){a}]\rbrace\ .
\end{eqnarray}  
Here ${H}$ is the full Hamiltonian, $\kappa$ is the photon loss rate of cavity and $\bar{n}$ is the average number of thermal photons.

We shall now investigate the gain that the two procedures (phase quick and dispersive interaction) bring to the dwell time of a qubit stored inside cavity $C$.

\subsection{Phase kick}

In the year 2002, Morigi et al. in \cite{morigi} have theoretically shown that the unitary evolution of a harmonic oscillator coupled to a two-level system can be undone by a suitable manipulation of the two-level system, specifically, by a quasi-instantaneous phase change (or commonly known as phase kick). This quasi-instantaneous phase kick enables us to isolate the dissipative evolution to which the oscillator may be exposed. Three years later, Meunier et al. in \cite{meunier} implemented this technique in CQED, showing that the collapse of a Rabi atomic oscillation in a coherent field is reversible, i.e., the phase kick applied to the atom induces revival phenomena at different instants of time than would occur spontaneously.

\begin{figure}[h]
	\centering
	\includegraphics[scale=0.5]{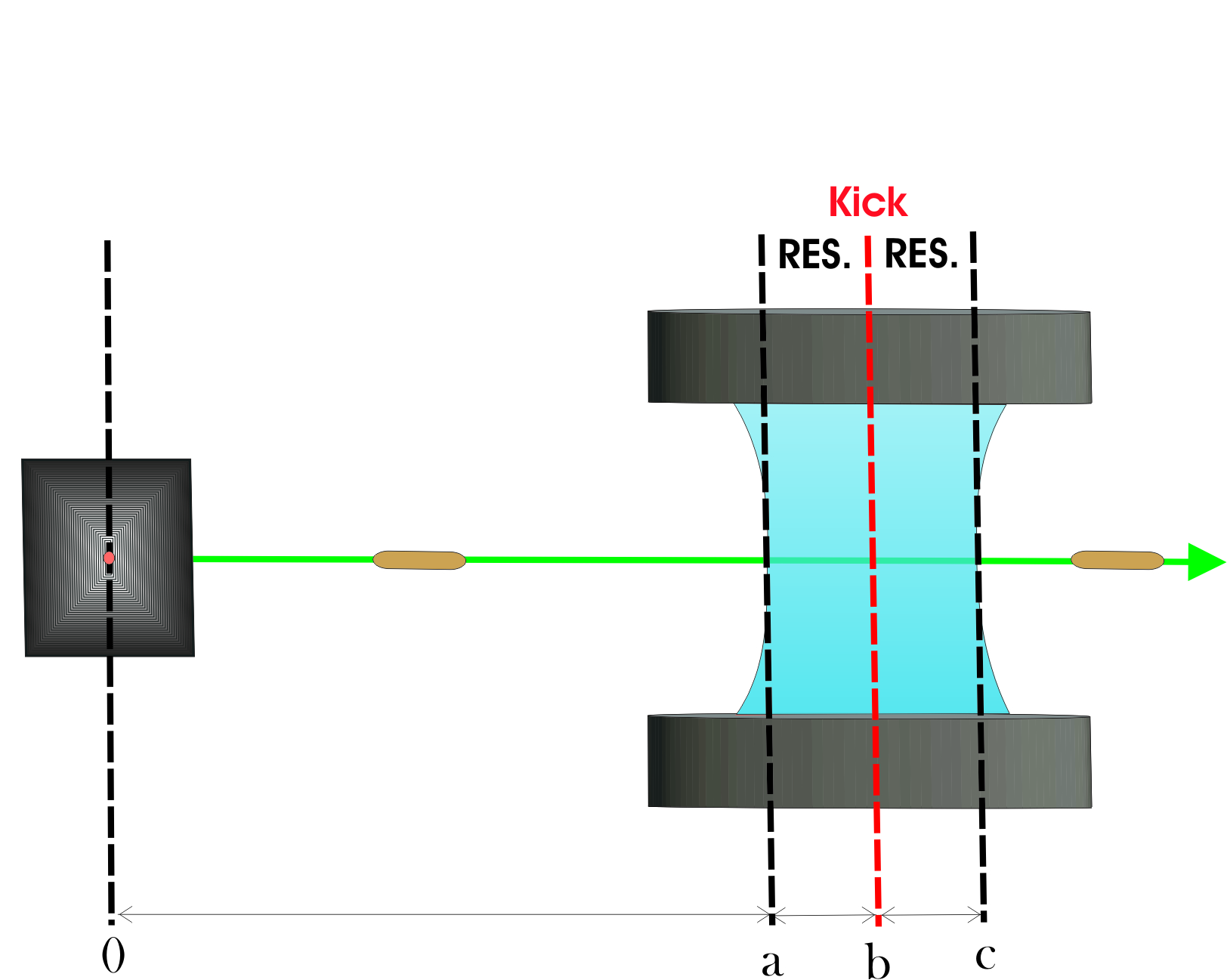}
	\caption{We have the general view, between the exit of the box B and the exit of the cavity $C$, of the path that will be traveled by each atom pertaining to the beam in the phase kick procedure.}
	\label{fig2}
\end{figure}

Therefore, we apply this method as a way of monitoring and preserving the coherence of a qubit stored in the cavity mode $C$. For this, consider a sample of atoms in the ground state, sent in the form of a beam, passing through the cavity with atomic velocity $v_0$ adjusted so that each atom interacts with the field of the interior of the cavity $C$ for a time $T_1+kick+T_2=T+kick$, undergoing three successive evolutions: i) the first resonant, for a time interval $T_1$; ii) the second is the quasi-instantaneous phase kick applied in the atom; iii) and the third again resonant, for a time interval $T_2$. As shown in Fig. \ref{fig2}.

Thus, each atom entering the cavity first will interact resonantly for a time interval $T_1=\pi/\Omega$, where $\Omega$ is the Rabi frequency modified by coupling with the reservoir, we say that the atom-field system has suffered a pulse $\pi$. With this, we map the field state in the atom. At the end of this time interval $T_1$ the atom undergoes a quasi-instantaneous phase kick corresponding to the unitary operation ${U}_{kick}=\left| e\right\rangle \left\langle e\right|-\left| g\right\rangle \left\langle g\right|={\sigma}_z$, the Hamiltonian generator of this phase kick is ${H}_{kick}=\hbar\pi{\sigma}_-{\sigma}_+\delta(t-T_1)$. Therefore, the effect of the kick after $T_1$ will be: ${\rho}(T_1)_{kick}={U}_{kick}{\rho}(T_1){U}^{-1}_{kick}={\sigma}_z{\rho}(T_1){\sigma}_z$, i.e., it will put a phase factor $e^{i\pi}$ in some of the coherence terms. It then, returns the resonant interaction for a time interval $T_2=\pi/\Omega$ and the atom-field system undergoes a new pulse $\pi$, returning the state to the cavity mode.

The first atom which initially is in the ground state $\left|g\right\rangle$, find the field inside of the cavity $C$ in the state ${\rho}_C(0)=\rho_{1,1}(0)\left| 1\right\rangle \left\langle 1\right| + [1-\rho_{1,1}(0)] \left| 0\right\rangle \left\langle 0\right| +[\rho_{1,0}(0)\left| 1\right\rangle \left\langle 0\right| +\mathrm{c.h.}]$, such that, after the interaction we can trace out the atomic state and get the field state. Between the exit of the first atom and the entrance of the second, there may be a window of time $t_1$ that the field inside $C$ evolves freely, subject to losses. In this way, after the passage of the $N$-th atom of the beam through the cavity, the field state becomes
\begin{multline}
{\rho}_C(NT+\mathcal{T})_{kick}=\rho_{1,1}(0)\prod_{i=1}^{N}\eta^2(\kappa,\Omega,T,t_i)\left| 1\right\rangle \left\langle  1\right|\\+[1-\rho_{1,1}(0)\prod_{i=1}^{N}\eta^2(\kappa,\Omega,T,t_i)]\left| 0\right\rangle \left\langle  0\right|\\+[\rho_{1,0}(0)\prod_{i=1}^{N}\eta(\kappa,\Omega,T,t_i)\left| 1\right\rangle \left\langle  0\right|+\mathrm{c.h.}]\ ,
\end{multline}
where $\mathcal{T}=\sum_{i=1}^{N-1}t_i$, and $\eta(\kappa,\Omega,T,t)=(1+2\frac{\kappa^2}{\Omega^2})e^{-\kappa(T+2t)/2}$. The dwell time of the population terms $T^{p}_r$ and the coherence terms $T^{c}_r$ of the state within the cavity are respectively
\begin{equation}
T_r^p=\frac{NT+\mathcal{T}}{\kappa(NT+2\mathcal{T})-2N\ \mathrm{ln}[1+2\frac{\kappa^2}{\Omega^2}]}
\end{equation}
and
\begin{equation}
T_r^c=\frac{2NT+2\mathcal{T}}{\kappa(NT+2\mathcal{T})-2N\ \mathrm{ln}[1+2\frac{\kappa^2}{\Omega^2}]}\ .
\end{equation}

In order avoid effects of collective coupling, the density $\lambda$ of the atomic beam is reduced to a value less than unity. Therefore, a beam with $N$ atoms will have on average $\lambda N$ atoms traversing the cavity and other $(1-\lambda)N$ atoms absent. It becomes relevant to examine the effects that absent atoms exert on the experimental procedure. This way, we get the following new dwell time
\begin{equation}
T_r^c=\frac{1}{\kappa}\frac{1}{1-\lambda\ \alpha(\kappa,\Omega,T)}
\end{equation}
and
\begin{equation}
T_r^p=\frac{1}{2\kappa}\frac{1}{1-\lambda\ \alpha(\kappa,\Omega,T)}\ ,
\end{equation}
where $\alpha(\kappa,\Omega,T)=\frac{1}{2T}\{T+\frac{2}{\kappa}\mathrm{ln}[1+2\frac{\kappa^2}{\Omega^2}]\}$. Note that, the term $1/[1-\lambda\ \alpha(\kappa,\Omega,T)]$ is responsible for improving the dwell time of the cavity. Thus, we can define a gain function $g(\kappa,\Omega,T,\lambda)$, which will quantify the increase that the dwell time obtained in this experimental proposal
\begin{equation}\label{funcao de ganho kick 3d}
	g(\kappa,\Omega,T,\lambda)=\bigg\{\frac{1}{1-\lambda\alpha(\kappa,\Omega,T)}-1\bigg\} \ .
\end{equation}
\begin{figure}[!htb]
	\centering
	\includegraphics[scale=.4]{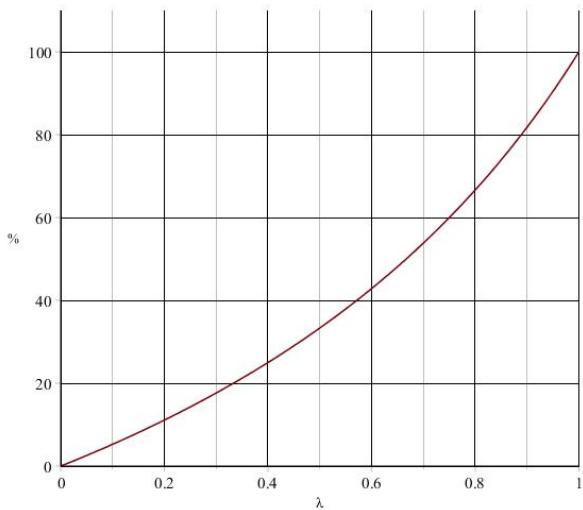}
	\caption{Gain of the dwell time of the cavity $g(\kappa,\Omega,T,\lambda)$ as a function of the linear density $\lambda$ of the atomic beam. In this graphic, we use realistic data taken from \cite{kuhr2007}.}
	\label{grafico ganho kick}
\end{figure}

The Fig. \ref{grafico ganho kick} shows that, when $\lambda\longrightarrow 1$, i.e., when there is always one atom inside the cavity, we can double the dwell time of the qubit inside the cavity. Another way of smoothing the loses and inhibiting the decoherence process of the qubit stored in the cavity, is to share the information (qubit) with an more stable auxiliary system.

\subsection{Dispersive interaction}

\begin{figure}[h]
	\centering
	\includegraphics[scale=0.5]{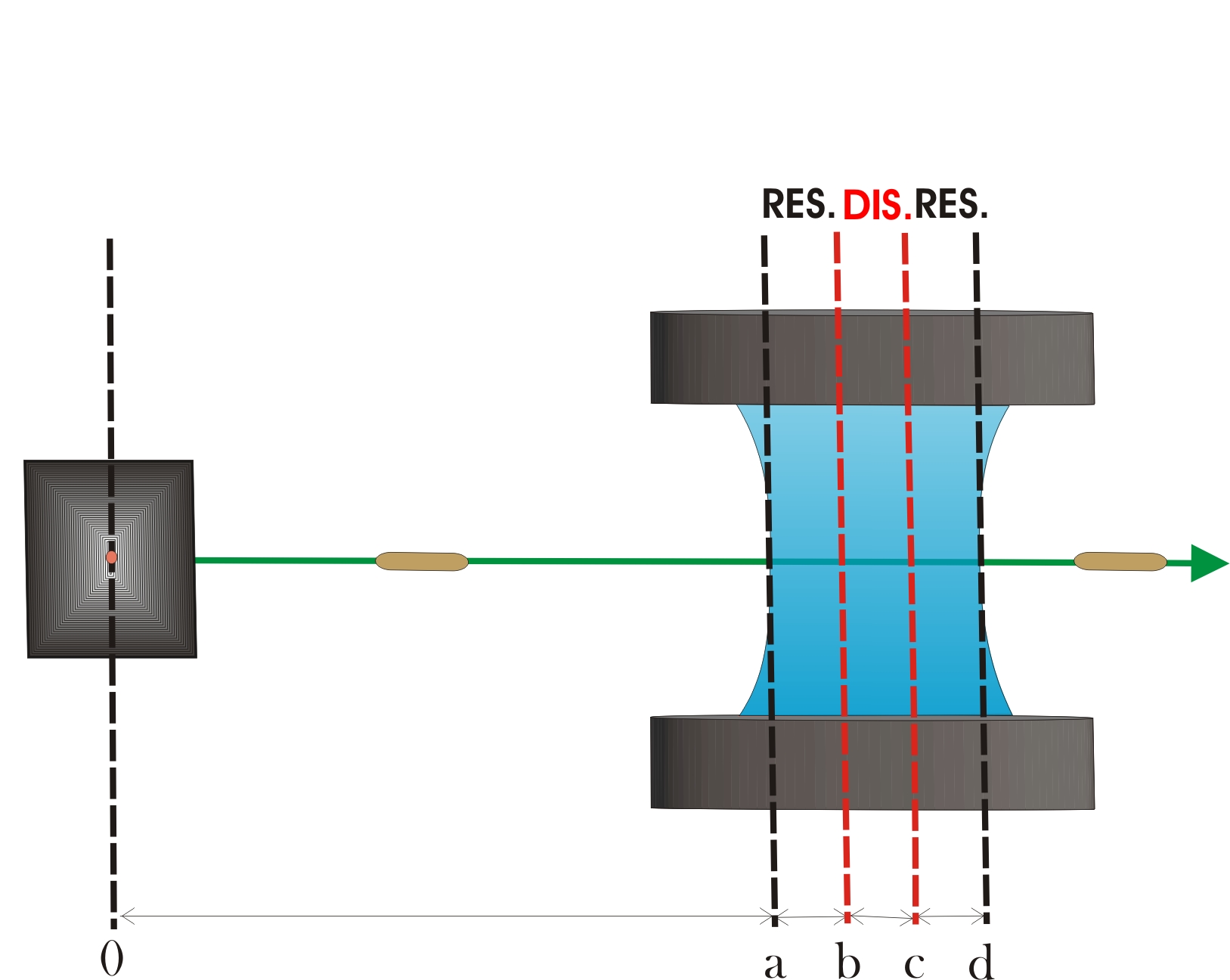}
	\caption{We have the general view, between the exit of the box B and the exit of the cavity $C$, of the path that will be traveled by each atom pertaining to the beam in the dispersive interaction procedure.}
	\label{fig4}
\end{figure}

Researchers of the Ecole Normale Supérieure in Paris showed in \cite{bertet} that by means of controlled interactions in a superconducting cavity $C$, is possible to experimentally investigate the principle of complementarity \cite{wheeler}. Using this idea of controlled interactions in $C$, we proposed the second method. For this, consider a sample of atoms in the ground state, sent in the form of a beam, transposing the cavity with atomic velocity $v_0$ adjusted so that each atom interacts with the field of the interior of the cavity $C$ for a time $T_1+\tau+T_2=T+\tau$, undergoing three successive evolutions: i) the first resonant, for a time interval $T_1$; ii) the second dispersive, for a time interval $\tau$; iii) and the third again resonant, for a time interval $T_2$. As shown in Fig. \ref{fig4}.

So, again each atom entering the cavity first will interact resonantly for a time interval $T_1=\pi/\Omega$, suffering a pulse $\pi$. With this, we map the field state in the atom. Soon after, we raise the resonant interaction via the Stark effect and let the system evolve dispersively over a period of time $\tau$, the dispersive interaction hides the atomic state of the effects of losses. After returns the resonant interaction for a time interval $T_2=\pi/\Omega$ and the atom-field system undergoes a new pulse $\pi$, returning the state to the cavity mode. It can be seen that the time that the system (atom+field) passes interacting resonantly is $T=2\pi/\Omega$, which will give us a phase factor $e^{i\pi}$ in the state $\left|g,1\right\rangle$, this same state during time $\tau$ accumulates a phase $e^{i\omega\tau}$ from the dispersive coupling. In order to reverse the state change caused by the phase factors, we can choose $\tau=\pi/\omega$.

Thus, again considering the initial state ${\rho}_C(0)$ for the field inside the cavity. After the passage of the N-th atom of the beam the field state becomes
\begin{multline}
{\rho}_C(NT+N\tau+\mathcal{T})=\rho_{1,1}(0)\prod_{i=1}^{N}\Gamma^2(\kappa,\Omega,\tau,T,t_i)\left| 1\right\rangle \left\langle  1\right| \\+[1-\rho_{1,1}(0)\prod_{i=1}^{N}\Gamma^2(\kappa,\Omega,\tau,T,t_i)]\left| 0\right\rangle \left\langle  0\right|\\+[\rho_{1,0}(0)\prod_{i=1}^{N}\Gamma(\kappa,\Omega,\tau,T,t_i)\left| 1\right\rangle \left\langle  0\right|+\mathrm{c.h.}]\ ,
\end{multline}
where $\Gamma(\kappa,\Omega,\tau,T,t)=[1+\frac{\kappa}{\Omega}f(\kappa,\Omega,\tau)]e^{-\kappa(T+2t)/2}$. While for this new procedure the dwell times of the population terms $T^{p}_r$ and the coherence terms $T^{c}_r$, taking into account the density of the atomic beam, are respectively
\begin{equation}
T_r^p=\frac{1}{2\kappa}\frac{1}{1-\lambda\ \varsigma(\kappa,\Omega,\tau,T)}
\end{equation}
and
\begin{equation}
T_r^c=\frac{1}{\kappa}\frac{1}{1-\lambda\ \varsigma(\kappa,\Omega,\tau,T)}\ ,
\end{equation}
here $\varsigma(\kappa,\Omega,\tau,T)=\frac{1}{T+\tau}\{\tau+\frac{T}{2}+\frac{1}{\kappa}\mathrm{ln}[1+\frac{\kappa}{\Omega}f(\kappa,\Omega,\tau)]\}$. We can define a new gain function $\mathcal{G}(\kappa,\Omega,\tau,T,\lambda)$, which will quantify the increase that the dwell time obtained in this experimental proposal
\begin{equation}\label{funcao.de.ganho}
\mathcal{G}(\kappa,\Omega,\tau,T,\lambda)=\bigg\{\frac{1}{1-\lambda\varsigma(\kappa,\Omega,\tau,T)}-1\bigg\} \ .
\end{equation}
\begin{figure}[!htb]
	\centering
	\includegraphics[scale=.33]{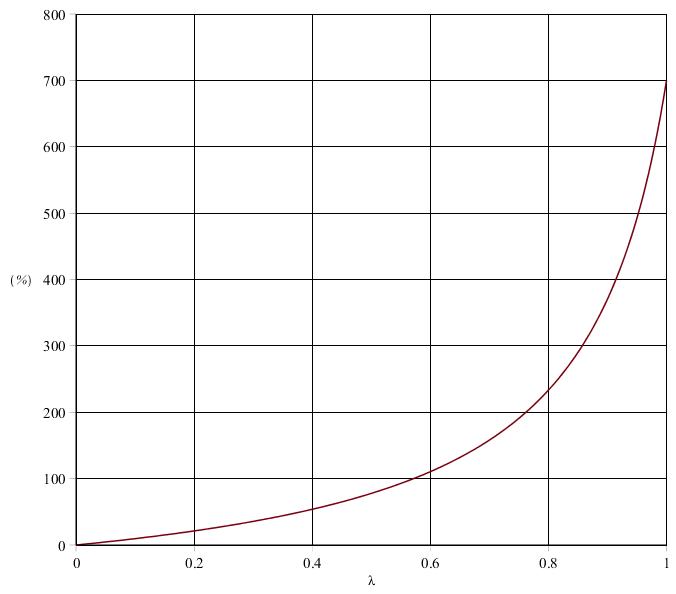}
	\caption{Gain of the dwell time of the cavity $g(\kappa,\Omega,T,\lambda)$ as a function of the linear density $\lambda$ of the atomic beam. In this graphic, we use realistic data taken from \cite{kuhr2007} and we consider the minimum detuning of $\delta=3G$, which provides $\tau=6\pi/\Omega$ (see \cite{brune45}).}
	\label{grafico.ganho}
\end{figure}

The Fig. \ref{grafico.ganho} shows that, when $\lambda\longrightarrow 1$, i.e., when there is always one atom inside the cavity, we can have a gain of up to 700 \% in this new experimental proposal. That is an expressive increase in the dwell time of the qubit inside the cavity.\\

\section{Dispersion of the atomic velocity}

In our experimental proposal the atom acts as a ancilla (auxiliary system), where the qubit is stored temporarily protecting it from the effects of decoherence. However, when interacting with the field, the atom can impose damages to the same to depend on the perfection of how the atomic state was prepared. The essential ingredient for a good control of the experimental sequence is the selection of the atomic velocity. The control of the atomic velocity is made very accurately, but there is always a random dispersion during the selection process that ranges from $0$ to $\pm 2$ m/s around the optimal velocity $v_0$ \cite{raimond},  making the interactions occur of imperfect ways. So, when we take into account the deviation of the velocity in our procedures we have:

\subsection{Phase kick}

In our first proposal, the total interaction time is $T\approx 1,96\times10^{-5}$ s. The effective length of atom-field interaction is $\sqrt{\pi}w_0\approx 1$ cm, where $w_0$ is the \textquotedblleft{waist}\textquotedblright of the transverse electromagnetic mode in the interior of $C$ and is of the order of 6mm \cite{raimond}. This gives us an optimal velocity of interaction $v_0\approx 510$ m/s. At this rate, the interaction of the atom with the field mode is perfect, so that, the two pulses $\pi$ occur completely. But as discussed above, small deviations can occur bringing us losses.

When we take into account the deviations of the atomic velocity, the $i$-th atom belonging to the beam that has velocity $v_i=v_0+\Delta v_i$ will have traveled a distance of $\mathrm{b}+\mathrm{b}\Delta v_i/v_0$ in the moment of the kick, consequently affecting the interactions time (see Fig. \ref{fig2}). We will have the following changes in our interaction times:
\begin{itemize}
	\item[i)] $T_1\longrightarrow T_1+ \dfrac{\mathrm{a}}{v_0}\dfrac{\Delta v_i}{v_0} = T_1 + \Delta t^i_{\pi_1}$\ ;  \item[ii)] $T_2 \longrightarrow T_2 - \dfrac{\mathrm{c}}{v_0}\dfrac{\Delta v_i}{v_0} = T_2 - \Delta t^{i}_{\pi_2}$\ .
\end{itemize}

Starting again from the initial state ${\rho}_C(0)$ to the field inside the cavity, but taking the new interaction times. After passing the $N$-th atom of the beam through of $C$, the new dwell times of the population terms $T_R^p$ and coherence terms $T_R^c$, taking into account the dispersion of the atomic velocity, are
\begin{widetext}
\begin{equation}\label{TR-diagonal kick}
T_R^p=\frac{1}{2\kappa}\frac{2T}{T-\frac{2}{\kappa}\mathrm{ln}\big[1+2\frac{\kappa^2}{\Omega^2}\big]+(1-\lambda)\big\{T+\frac{2}{\kappa}\mathrm{ln}\big[1+2\frac{\kappa^2}{\Omega^2}\big]\big\}+D(\kappa,\Omega,\mathrm{a},w_0,v_0)\left( \frac{\Delta v}{v_0}\right)^2}
\end{equation}
and
\begin{equation}\label{tempo.de.retencao.TR kick}
T_R^c=\frac{1}{\kappa}\frac{2T}{T-\frac{2}{\kappa}\mathrm{ln}\big[1+2\frac{\kappa^2}{\Omega^2}\big]+(1-\lambda)\big\{T+\frac{2}{\kappa}\mathrm{ln}\big[1+2\frac{\kappa^2}{\Omega^2}\big]\big\}+\big[D(\kappa,\Omega,\mathrm{a},w_0,v_0)+W(\kappa,\Omega,\mathrm{a},w_0,v_0)\big]\left( \frac{\Delta v}{v_0}\right)^2}\ ,
\end{equation}
with
\begin{eqnarray}\nonumber
D(\kappa,\Omega,\mathrm{a},w_0,v_0)= \frac{3}{4}\kappa\frac{[\mathrm{a}+(\mathrm{a}+\sqrt{\pi}w_0)]^2}{v_0^2}+\frac{1}{4}\frac{\Omega^2}{\kappa}\frac{[\mathrm{a}+(\mathrm{a}+\sqrt{\pi}w_0)]^2}{v_0^2}\ ,
\end{eqnarray}
\begin{eqnarray}\nonumber
W(\kappa,\Omega,\mathrm{a},w_0,v_0)=-\frac{1}{4}\kappa\frac{(\mathrm{a}^2+(\mathrm{a}+\sqrt{\pi}w_0)^2)}{v_0^2}-\frac{3}{2}\kappa\frac{\mathrm{a}(\mathrm{a}+\sqrt{\pi}w_0)}{v_0^2}\ .
\end{eqnarray}
\end{widetext}

It is interesting to note that, the denominator of the dwell times independently brings the terms that carry the information corresponding to the atomic presence, the absent atom and the dispersion of the atomic velocity, respectively. For quantum information purposes, we must preserve the coherence term of our state for as long as possible. Therefore
\begin{multline}\nonumber
\lambda\bigg(T+\frac{2}{\kappa}\mathrm{ln}\Big[1+2\frac{\kappa^2}{\Omega^2}\Big]\bigg)>\big[D(\kappa,\Omega,\mathrm{a},w_0,v_0)\\+W(\kappa,\Omega,\mathrm{a},w_0,v_0)\big]\left( \frac{\Delta v}{v_0}\right)^2\ ,
\end{multline}
which leads us to the conclusion that, in order to obtain significant gains in dwell time, the following inequality must be satisfied
\begin{equation}
1>\frac{\pi}{2\lambda}\bigg(1+\frac{2\Omega}{\pi}\frac{\mathrm{a}}{v_0}\bigg)\bigg(\frac{\Omega}{\kappa}\bigg)\bigg(\frac{\Delta v}{v_0}\bigg)^2
\end{equation}
or, using realistic data extracted from \cite{kuhr2007, brune45}
\begin{equation}
10^{-3}>\bigg(\frac{\Omega}{\kappa}\bigg)\bigg(\frac{\Delta v}{v_0}\bigg)^2\ .
\end{equation}

This equation indicates to us that only the improvement of our cavity $C$ is not enough to obtain an expressive increase of our dwell time $T_R^c$, i.e., if there is no improvement in the selection of the atomic velocity, decreasing its standard deviation $\Delta v$, we will not have significant gains in the dwell time of the qubit inside the cavity. This unprecedented result establishes bonds for the existence of gain in the dwell time of the qubit inside the cavity.

To better understand how influential the dispersion of the atomic velocity may be in the CQED experiments, we define a new gain function $g'(\kappa,\Omega,T,\lambda,\Delta v)$ for our dwell time as a function of $\lambda$ and $\Delta v$
\begin{equation}\label{funcao.de.ganho2 kick}
g'(\kappa,\Omega,T,\lambda,\Delta v)=\{\kappa T_R^c-1\} \ .
\end{equation}

\begin{figure}[!htb]
	\centering
	\includegraphics[scale=.33]{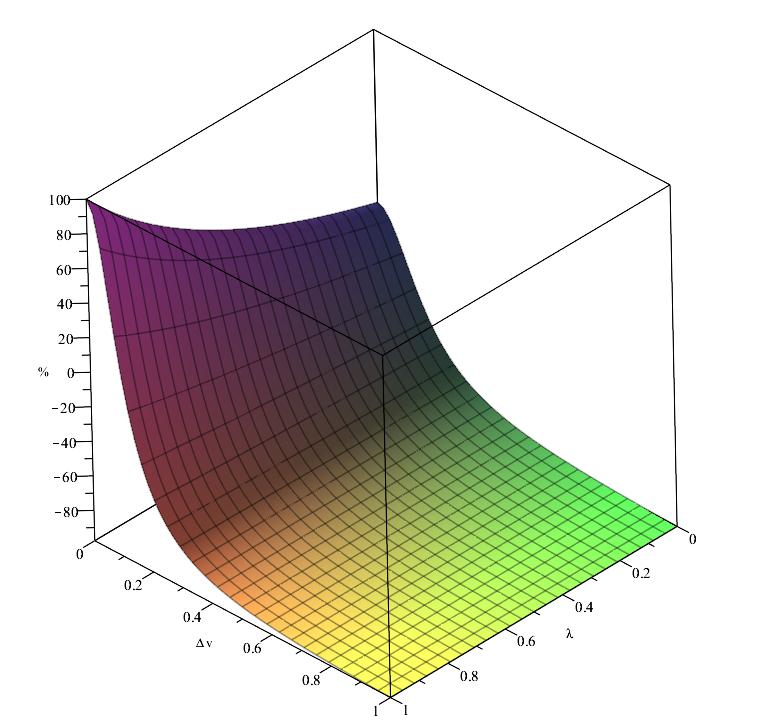}
	\caption{Gain of the dwell time of the cavity $g'(\kappa,\Omega,T,\lambda,\Delta v)$ as a function of the linear density $\lambda$ and the standard deviation of the atomic velocity $\Delta v$ of the beam. In this graphic, we use realistic data taken from \cite{kuhr2007, brune45}.}
	\label{grafico.ganho2_kick}
\end{figure}

In the graph of the gain function Fig. \ref{grafico.ganho2_kick}, we can observe that, even when the linear density $\lambda$ of the atomic beam tends to one, i.e., when there is always an atom inside the cavity. We will only double our dwell time when the standard deviation $\Delta v$ in the selection of the atomic velocity tends to zero.

At the end of the passage of the $N$-th atom of the beam through the cavity, the field state in its inside taking into account the dispersion of the atomic velocity will be
\begin{multline}\label{estado final kick}
{\rho}_C(\Delta t)_{kick}=\rho_{1,1}(0)e^{-2\bar{\kappa}\Delta t}\left|1\right\rangle \left\langle 1\right|\\+\big[1-\rho_{1,1}(0)e^{-2\bar{\kappa}\Delta t}\big]\left|0\right\rangle \left\langle 0\right|\\+\big[\rho_{1,0}(0)e^{-\bar{\kappa}\Delta t}e^{-\bar{F}\Delta t}\left|1\right\rangle \left\langle 0\right|+ \mathrm{c.h.}\big]\ ,
\end{multline}
where $\Delta t$ is the total time of field evolution,
\begin{eqnarray*}
\bar{\kappa}&=&\frac{1}{2T_R^p}\ ,\\
\bar{F}&=&\kappa.\frac{W(\kappa,\Omega,\mathrm{a},w_0,v_0)\left(\frac{\Delta v}{v_0}\right)^2}{2T}\approx 10^{-1}\left( \frac{\Delta v}{v_0}\right)^2\ .
\end{eqnarray*}

\subsection{Dispersive interaction}

For our second proposal, the total interaction time is $4T\approx 7,85\times10^{-5}$ s, whereas the minimal detuning is $\delta=3G$ \cite{brune45}. As the effective length of atom-field interaction is $\sqrt{\pi}w_0\approx 1$ cm, the optimal velocity of interaction is $v_0\approx 127,5$ m/s. At this rate, the interaction of the atom with the field mode is perfect, so that, both the pulses $\pi$ and the dispersive interaction are complete. However, as already discussed, small deviations can occur in the atomic selection process and this implies losses in the field state.

When we take into account the deviations of the atomic velocity, the $i$-th atom belonging to the beam that has velocity $v_i=v_0+\Delta v_i$ will have traveled a distance of $\mathrm{b}+\mathrm{b}\Delta v_i/v_0$ at the moment that the resonant interaction is suspended, resulting in an imperfect pulse $\pi$ (see Fig. \ref{fig4}). Note that, the deviations in the atomic velocity are very small, and the inequality $|\Delta v_i|\ll v_0$ will be valid. Since the phase shifts caused by the dispersion of the atomic velocity during the dispersive interaction will be very small, they can be neglected, i.e., the interaction time $\tau$ between $\mathrm{b}$ e $\mathrm{c}$ remains unchanged. On the other hand, the time of interaction between $\mathrm{c}$ and $\mathrm{d}$ is influenced by the velocity deviation. We will have the following changes in our interaction times:
\begin{itemize}
	\item[i)] $T_1\longrightarrow T_1+ \dfrac{\mathrm{a}}{v_0}\dfrac{\Delta v_i}{v_0} = T_1 + \Delta t^i_{\pi_1}$\ ;  \item[ii)] $T_2 \longrightarrow T_2 - \dfrac{\mathrm{d}}{v_0}\dfrac{\Delta v_i}{v_0} = T_2 - \Delta t^{i}_{\pi_2}$\ .
\end{itemize}

Starting again from the initial state ${\rho}_C(0)$ to the field inside the cavity, but taking the new interaction times. After passing the $N$-th atom of the beam through of $C$, the new dwell times of the population terms $T_R^p$ and coherence terms $T_R^c$, taking into account the dispersion of the atomic velocity, are
\begin{widetext}
\begin{equation}\label{TR-diagonal}
T_R^p=\frac{1}{2\kappa}\frac{2(T+\tau)}{T-\frac{2}{\kappa}\mathrm{ln}\big[1+\frac{\kappa}{\Omega}f(\kappa,\Omega,\tau)\big]+(1-\lambda)\{T+2\tau+\frac{2}{\kappa}\mathrm{ln}[1+\frac{\kappa}{\Omega}f(\kappa,\Omega,\tau)]\}+\Lambda(\kappa,\Omega,\tau,\mathrm{a},\omega_0,v_0)\left( \frac{\Delta v}{v_0}\right)^2}
\end{equation}
and
\begin{multline}\label{tempo.de.retencao.TR}
T_R^c=\frac{1}{\kappa}\frac{2(T+\tau)}{T-\frac{2}{\kappa}\mathrm{ln}\big[1+\frac{\kappa}{\Omega}f(\kappa,\Omega,\tau)\big]+(1-\lambda)\{T+2\tau+\frac{2}{\kappa}\mathrm{ln}[1+\frac{\kappa}{\Omega}f(\kappa,\Omega,\tau)]\}+\big[\Lambda(\kappa,\Omega,\tau,\mathrm{a},\omega_0,v_0)}\\\frac{}{+\xi(\kappa,\Omega,\tau,\mathrm{a},\omega_0,v_0)\big]\left( \frac{\Delta v}{v_0}\right)^2}\ ,
\end{multline}
with
\begin{eqnarray}\nonumber
\Lambda(\kappa,\Omega,\tau,\mathrm{a},\omega_0,v_0)&=&	\frac{1}{\Omega}f(\kappa,\Omega,\tau)\bigg\{\frac{\Omega^2}{4}\bigg[\frac{\mathrm{a}^2+(\mathrm{a}+\sqrt{\pi}\omega_0)^2}{v_0^2}\bigg]+\frac{\Omega^4}{2\kappa^2}\frac{\mathrm{a}(\mathrm{a}+\sqrt{\pi}\omega_0)}{v_0^2}\bigg\}+\frac{\Omega^2}{4\kappa}\left( \frac{\sqrt{\pi}\omega_0}{v_0}\right)^2\\ \nonumber
& &+\frac{\Omega}{2}f(\kappa,\Omega,\tau)\bigg[\frac{\mathrm{a}^2+(\mathrm{a}+\sqrt{\pi}\omega_0)^2}{2v_0^2}-\frac{\mathrm{a}(\mathrm{a}+\sqrt{\pi}\omega_0)}{v_0^2}e^{-\kappa\tau}\bigg]-\frac{\kappa}{4}\bigg(\frac{\sqrt{\pi}\omega_0}{v_0}\bigg)^2e^{-2\kappa\tau}\ ,
\end{eqnarray}
\begin{eqnarray}\nonumber
\xi(\kappa,\Omega,\tau,\mathrm{a},\omega_0,v_0)=-\frac{\Omega}{2}f(\kappa,\Omega,\tau)\bigg[\frac{\mathrm{a}^2+(\mathrm{a}+\sqrt{\pi}\omega_0)^2}{2v_0^2}-\frac{\mathrm{a}(\mathrm{a}+\sqrt{\pi}\omega_0)}{v_0^2}e^{-\kappa\tau}\bigg]+\frac{\kappa}{4}\bigg(\frac{\sqrt{\pi}\omega_0}{v_0}\bigg)^2e^{-2\kappa\tau}\ .
\end{eqnarray}
\end{widetext}

To obtain some gain in the dwell time of the cavity is has to that
\begin{multline}\nonumber
\lambda\bigg(T+2\tau+\frac{2}{\kappa}\mathrm{ln}\Big[1+\frac{\kappa}{\Omega}f(\kappa,\Omega,\tau)\Big]\bigg)>\\\big[\Lambda(\kappa,\Omega,\tau,\mathrm{a},\omega_0,v_0)+\xi(\kappa,\Omega,\tau,\mathrm{a},\omega_0,v_0)\big]\left( \frac{\Delta v}{v_0}\right)^2\ .
\end{multline}
which leads us to conclude that, the following inequality must be satisfied
\begin{eqnarray}
1>\frac{8\pi}{7\lambda}\bigg(1+\frac{3\kappa}{2}\frac{\mathrm{a}}{v_0}\bigg)\bigg(\frac{\Omega}{\kappa}\bigg)\bigg(\frac{\Delta v}{v_0}\bigg)^2
\end{eqnarray}
or, using realistic data extracted from \cite{kuhr2007, brune45}
\begin{equation}
	10^{-1}>\bigg(\frac{\Omega}{\kappa}\bigg)\bigg(\frac{\Delta v}{v_0}\bigg)^2\ .
\end{equation}

This equation confirms and reinforces the result obtained in the previous proposal, pointing out that, only the improvement of the cavity $C$ is not enough to obtain an expressive increase of our dwell time $T_R^c$, i.e., if there is no improvement in selection of the atomic velocity, decreasing its standard deviation $\Delta v$, we will not have significant gains in the dwell time of the qubit inside the cavity.

Defining a new gain function $\mathcal{G}'(\kappa,\Omega,\tau,T,\lambda,\Delta v)$ for our dwell time as a function of $\lambda$ and $\Delta v$, we have
\begin{equation}\label{funcao.de.ganho2}
\mathcal{G}'(\kappa,\Omega,\tau,T,\lambda,\Delta v)=\{\kappa T_R^c-1\} \ .
\end{equation}

\begin{figure}[!htb]
	\centering
	\includegraphics[scale=.3]{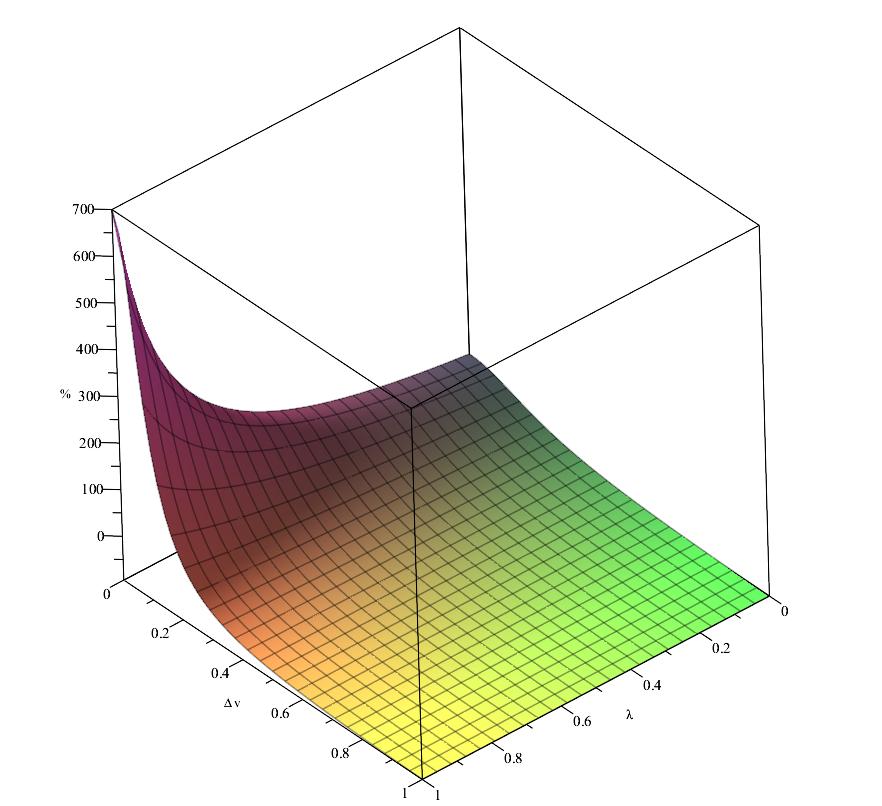}
	\label{grafico.ganho2}
	\caption{Gain of the dwell time of the cavity $\mathcal{G}'(\kappa,\Omega,\tau,T,\lambda,\Delta v)$ as a function of the linear density $\lambda$ and the standard deviation of the atomic velocity $\Delta v$ of the beam. In this graphic, we use realistic data taken from \cite{kuhr2007, brune45}.}
\end{figure}

At the end of the passage of the $N$-th atom of the beam through the cavity, the field state in its inside taking into account the dispersion of the atomic velocity will be
\begin{multline}\label{estado final}
{\rho}_C(\Delta t)=\rho_{1,1}(0)e^{-2\bar{\kappa}'\Delta t}\left|1\right\rangle \left\langle 1\right|\\+\big[1-\rho_{1,1}(0)e^{-2\bar{\kappa}'\Delta t}\big]\left|0\right\rangle \left\langle 0\right|\\+\big[\rho_{1,0}(0)e^{-\bar{\kappa}'\Delta t}e^{-\bar{F}'\Delta t}\left|1\right\rangle \left\langle 0\right|+ \mathrm{c.h.}\big]\ ,
\end{multline}
where $\Delta t$ is the total time of field evolution,
\begin{eqnarray*}
	\bar{\kappa}'&=&\frac{1}{2T_R^p}\ ,\\
	\bar{F}'&=&\kappa\frac{\xi(\kappa,\Omega,\tau,\mathrm{a},\omega_0,v_0)\left(\frac{\Delta v}{v_0}\right)^2}{2(T+\tau)}\approx 10^{-4}\left( \frac{\Delta v}{v_0}\right)^2\ .
\end{eqnarray*}

\section{Damping phase channel}

In both procedures presented, in the instant that the two-level atom couples with the field of the interior of $C$ and collects the information (qubit) stored therein, it decreases the loss of coherence caused by damping of the field on the qubit. However, as we saw, the atom at the same time can impose new decoherence processes depending on how well it was prepared. In a more complete treatment, we can make evolution of the initial state ${\rho}_C(0)$ by means of a master equation with damping amplitude and damping phase channels, of the type \cite{reservatorioFase}
\begin{eqnarray}\nonumber
\frac{d}{dt}{\rho}(t)&=&\beta\big\{\big[{a}{\rho}(t),{a}^{\dagger}\big]+\big[{a},{\rho}(t){a}^{\dagger}\big]\big\}+\varepsilon\big\{\big[{a}^{\dagger}{a}{\rho}(t),{a}^{\dagger}{a}\big]\\\label{eq mestra fase e amplitude}
& &+\big[{a}^{\dagger}{a},{\rho}(t){a}^{\dagger}{a}\big]\big\}\ ,
\end{eqnarray}
where the number of thermal photons $\bar{n}\approx 0$, while $\beta$ and $\varepsilon$ are the damping amplitude and damping phase channels constants, respectively. After a certain time interval $t$, the field is found in the state
\begin{eqnarray}\nonumber
{\rho}(t)&=&\rho_{1,1}(0)e^{-2\beta t}\left|1\right\rangle\left\langle 1\right|+\Big[1-\rho_{1,1}(0)e^{-2\beta t}\Big]\left|0\right\rangle\left\langle 0\right|\\\label{estado-evoluido}
& &+\Big[\rho_{1,0}(0)e^{-\beta t}e^{-\varepsilon t}\left|1\right\rangle\left\langle 0\right|+\mathrm{c.h.}\Big]\ .
\end{eqnarray}

A succinct analysis of the above equation, shows that, the damping phase channel acts on the coherence terms of the density operator of the system of interest ${\rho}(t)$, by means of the factor $e^{-\varepsilon t}$. Thus, by comparing the final state of the evolved density operator by means of a master equation with damping amplitude and damping phase (\ref{estado-evoluido}) with the final states of the field inside the cavity $C$ at the end of the passing of the $N$-th beam atom (\ref{estado final kick}) and (\ref{estado final}), it can be observed that the dispersion of the atomic velocity is also responsible for the emergence of a damping phase channel, which induces loss of coherence on the field state inside the cavity through the factors $e^{-\bar{F}\Delta t}$ and $e^{-\bar{F}'\Delta t}$ for the first and second procedure, respectively. However, as $\bar{F}$ and $\bar{F}'$ are $\ll 1$, the action of the phase reservoir is very weak when compared to that of decay, making it negligible in typical cavity experiments.

\section{Fidelity}

One way of quantifying how close the evolved field state is from the initial state, i.e., the amount of information lost after the passage of the $N$-th atom of the beam, is by means of a measure called fidelity \cite{artigofidelity}. Let us assume an initial state of the type $\left|\psi_1\right\rangle =[\left|0\right\rangle +e^{i\phi}\left|1\right\rangle]/\sqrt{2}$, that corresponds to a particular case of the arbitrary initial state ${\rho}_C(0)$. The initial state $\left|\psi_1\right\rangle$ evolves over the action of a reservoir with damping amplitude and damping phase channels, which as we saw in the previous section, re-creates the dynamics of preservation of a qubit within $C$. Therefore, at the end of evolution, the state becomes	
\begin{eqnarray}\nonumber
\rho_2&=&\frac{1}{2}e^{-2\beta t}\left|1\right\rangle\left\langle 1\right|+\Big[1-\frac{1}{2}e^{-2\beta t}\Big]\left|0\right\rangle\left\langle 0\right|\\
& &+\Big[\frac{e^{i\phi}}{2}e^{-\beta t}e^{-\varepsilon t}\left|1\right\rangle\left\langle 0\right|+\mathrm{c.h.}\Big]\ .
\end{eqnarray}

In this way, fidelity to our system of interest is
\begin{equation}\label{fidelidade sistema}
F(\left|\psi_1\right\rangle\left\langle\psi_1\right|,\rho_2)=\left\langle\psi_1\right|\rho_2\left|\psi_1\right\rangle=\frac{1}{2}\big(1+e^{-\beta t}e^{-\varepsilon t})\ .
\end{equation}

\begin{figure}[!htb]
	\centering
	\includegraphics[scale=.37]{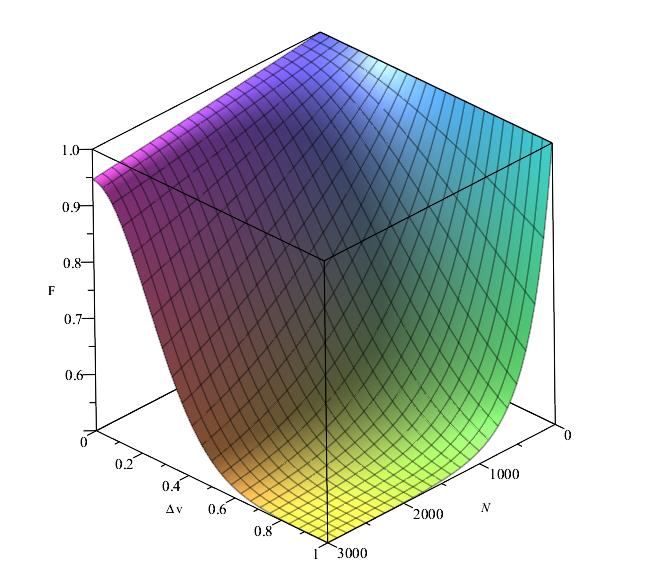}
	\caption{Fidelity $F(\left|\psi_1\right\rangle\left\langle\psi_1\right|,\rho_2)$ of the field state within the cavity as a function of the standard deviation of the atomic velocity $\Delta v$ and of the number of atoms $N$ belonging to the atomic beam that transposes the cavity $C$. In this graphic, we use the realistic data taken from \cite{kuhr2007}. It is worth remembering that, the total time of the qubit inside $C$ is $t=N\times1,96\times10^{-5}$s, where $1,96\times10^{-5}$s is the total interaction time for each atom and we consider the linear density of the beam $\lambda=1$.}
	\label{grafico fidelidade kick}
\end{figure}

For our first experimental proposal, in which a quasi-instantaneous phase kick is used as a way to monitor and preserve the coherence of a qubit stored inside the cavity $C$, the damping constants $\beta$ and $\varepsilon$ corresponds to the respective dynamic quantities $\bar{\kappa}$ and $\bar{F}$ state (\ref{estado final kick}). In this way, the fidelity in this case will have the behavior of the Fig. \ref{grafico fidelidade kick}, showing that, while the passage of 3000 atoms through the cavity leads to a loss of minimum fidelity, the increase in atomic velocity dispersion leads to an exponential loss of fidelity. This shows how influential the atomic velocity selection is.

\begin{figure}[!htb]
	\centering
	\includegraphics[scale=.35]{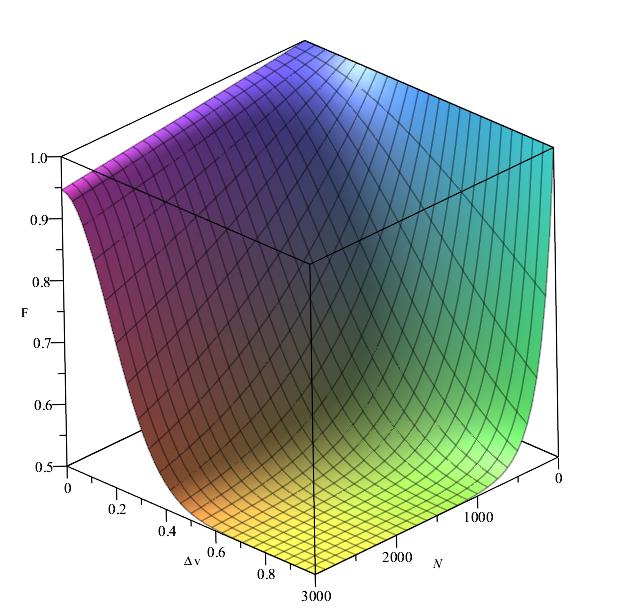}
	\caption{Fidelity $F(\left|\psi_1\right\rangle\left\langle\psi_1\right|,\rho_2)$ of the field state within the cavity as a function of the standard deviation of the atomic velocity $\Delta v$ and of the number of atoms $N$ belonging to the atomic beam that transposes the cavity $C$. In this graphic, we use the realistic data taken from \cite{kuhr2007}. It is worth remembering that, the total time of the qubit inside $C$ is $t=N\times7,85\times10^{-5}$s, where $7,85\times10^{-5}$s is the total interaction time for each atom, we consider the minimum detuning of $\delta=3G$, which provides $\tau=6\pi/\Omega$, and we consider the linear density of the beam $\lambda=1$.}
	\label{grafico fidelidade dispersivo}
\end{figure}

For the second experimental proposal, where we use controlled resonant and dispersive interactions as a way to preserve and monitor the qubit inside the cavity $C$, the damping constants $\beta$ and $\varepsilon$ correspond to the respective quantities dynamic $\bar{\kappa}'$ and $\bar{F}'$ of the state (\ref{estado final}). The fidelity to this case will have the following behavior of the Fig. \ref{grafico fidelidade dispersivo}, which, like the previous case, reaffirms our conclusions on the damages imposed by the dispersion of the atomic velocity.

\section{conclusion}

In short, in this work we analyze the damage imposed by the atom on the field state inside lossy superconducting cavities. To access such effects, we propose two procedures (phase kick and dispersive interaction) that partially preserve a qubit of the effects of losses of an imperfect quantum memory. We immediately showed that, in both procedures the dwell time of the qubit inside the cavity increases, being expressively superior to the second. From our accounts arises naturally, for each experimental proposal, a relation between the perfection of how the atomic state is prepared and the quality factor of the cavity. This result is unprecedented, and sets out the rules that points to an improvement in selection of the atomic velocity, so that we can have expressive gains in the dwell time of the qubit inside the cavity $C$. It was observed that at the instant that the atom-field coupling occurs, the atom can impose new noises to the field state inside the cavity, due to the dispersion of the atomic velocity. This new noises are similar to the action of a phase reservoir on the field state. However, the action of the phase reservoir becomes negligible when compared to the action of the decay reservoir in typical cavity experiments.
\newpage
\bibliography{bibliografia}

\end{document}